\documentclass[trackchanges, twocolumn]{aastex7}

\newcommand\aastex{AAS\TeX}

\newcommand{\msun}{\mbox{M$_{\odot}$}}

\newcommand{\mgii}{\mbox{Mg\,{\sc ii}}}

\newcommand{\feii}{\mbox{Fe\,{\sc ii}}}

\newcommand{\civ}{\mbox{C\,{\sc iv}}}

\newcommand{\nv}{\mbox{N\,{\sc v}}}

\newcommand{\siiv}{\mbox{Si\,{\sc iv}}}

\def\h2{$\rm H_2$}
\def\Nh2{$N$(H${_2}$)}

\def\kms{km\,s$^{-1}$}

\def\zem{$z_{\rm em}$}

\def\21{21-cm}

\def\t0{T$_{0}$}

\def\l{$\lambda$}

\def\c21{$C_{21}$}

\def\nv{N~{\sc v}}
\def\civ{C~{\sc iv}}
\def\siiv{Si~{\sc iv}}
\def\siv{S~{\sc iv}}
\def\aliii{Al~{\sc iii}}
\def\pv{P~{\sc v}}

\def\J11{J$1156+0856$}

\begin{document}

\title{Template \aastex v7 Article with Examples\footnote{Footnotes can be added to titles}}

\title{Transient LoBALs at high velocities: A Case of Extreme Broad Absorption Line Variability in J115636.82+085628.9   \footnote{Based on observations collected at Southern African Large Telescope (SALT; Programme IDs 2015-1-SCI-005, 2018-1-SCI-009, 2019-1-SCI-019 and 2020-1-SCI-011) and the European Organisation for Astronomical Research in the Southern Hemisphere under ESO programme 093.A-0255.}}

\author[orcid=0009-0001-2178-4022,sname='Aromal',gname=Pathayappura]{P. Aromal}
\affiliation{Physics and Astronomy department, University of Western Ontario, 1151 Richmond Street, London, N6A 3K7, Ontario, Canada}
\affiliation{Institute for Earth and Space Exploration, Western University, 1151 Richmond St., London, ON N6A 3K7, Canada}
\email[show]{apathaya@uwo.ca}  

\author[orcid=0000-0002-9062-1921,sname='Srianand',gname=Raghunathan]{R. Srianand}
\affiliation{IUCAA, Postbag 4, Ganeshkind, Pune 411007, India}
\email{anand@iucaa.in}

\author[orcid=0000-0001-6217-8101,sname='Gallagher',gname=Sarah]{S. C. Gallagher}
\affiliation{Physics and Astronomy department, University of Western Ontario, 1151 Richmond Street, London, N6A 3K7, Ontario, Canada}
\affiliation{Institute for Earth and Space Exploration, Western University, 1151 Richmond St., London, ON N6A 3K7, Canada}
\email{sgalla4@uwo.ca}  

\author[orcid=0000-0001-5937-331X,sname='Vivek',gname=M]{M. Vivek}
\affiliation{Indian Institute of Astrophysics, Bangalore 560034, India}
\email{vivek.m@iiap.res.in}

\author[sname='Petitjean',gname=Patrick]{P. Petitjean}
\affiliation{Institut d’Astrophysique de Paris, Sorbonne Universit\'e and CNRS, 98bis boulevard Arago, 75014 Paris, France}
\email{ppetitje@iap.fr}


\begin{abstract}
We present a multi-epoch spectroscopic study of the broad absorption line (BAL) quasar J115636.82+085628.9 ($z_{\mathrm{em}} = 2.1077$), based on five spectra spanning nearly two decades in the observer’s frame. 
This source exhibits remarkable variability in both low-ionization (LoBAL: \aliii\ and \mgii) and high-ionization (HiBAL: \civ\ and \siiv) absorption features. 
For the first time, we detect the emergence and subsequent disappearance of LoBAL troughs at high velocities ($\sim$20,000 \kms), coinciding with the strengthening and weakening of the corresponding HiBAL absorption.
The \civ\ BAL profile extends from $\sim$6,700 \kms\ to a conservative upper limit of 30,000 \kms\ and is composed of narrow, variable absorption features embedded within a broad, smooth envelope. 
Both \civ\ and \siiv\ BAL troughs exhibit dramatic equivalent width (EW) changes—among the most extreme reported to date. 
Notably, these EW variations are strongly anti-correlated with continuum flux changes inferred from optical photometric light curves. 
We interpret this variability as the result of a new absorbing flow transiting into our line of sight, increasing the shielding of a more distant, pre-existing outflow and giving rise to transient LoBAL absorption.
This scenario supports a unified picture in which LoBAL and HiBAL features arise from similar outflow structures, with observed differences governed primarily by line-of-sight column densities consistent with previous findings.
\end{abstract}




\section{Introduction}

Strong outflows from quasars manifest themselves as broad absorption lines (BALs) 
having widths of several 1000 \kms\ and outflow velocities reaching up to several 10,000 \kms\ with respect to the systemic redshift (\zem) of the quasars. 
About 10-20$\%$ of all optically selected quasars show BALs
\citep[][]{Weymann1991}
and this fraction can reach 40$\%$ if dust and other observational biases are properly taken into account \citep[][]{dai2008,allen2011}.
These outflows are believed to be related to accretion disk winds \citep[e.g.][]{murray1995} and have sufficient mechanical energy and momentum to provide significant feedback to the gas in galaxies \citep[e.g.][]{Arav2018}.
Also, recent studies provide growing evidence that BAL and non-BAL quasars exhibit similar distributions in emission-line outflow and physical properties, suggesting that they are drawn from the same parent population \citep{rankine2020, temple2024, Harum2024}.
%

BAL quasars are mainly divided into two main subtypes: (1) HiBAL sources that show absorption produced by highly ionized species such as \civ, \siiv, \nv\ etc. and (2) LoBAL sources where we see absorption from low-ionized species such as \aliii, \feii\ and \mgii\ etc.
The time variability in both HiBALs and LoBALs serves as a crucial tool for probing the nature, origin, and evolution of quasar outflows \citep{Srianand2001,Gibson2008, Gibson2010, filiz2012, capellupo2011, Vivek2014, cicco2018,lu2020,aromal2022,aromal2024}. Time-resolved studies of BAL profiles, both in LoBALs and HiBALs, have provided significant insights into the physical conditions, ionization states, covering fractions, and spatial locations of these outflows relative to the central source \citep[for example, see][]{Capellupo2012, Arav2018, green2023}. 
Monitoring BAL variability across different timescales provides stringent constraints on the lifetime and dynamics of the outflows, shedding light on the underlying physical mechanisms driving them. In particular, extreme cases of variability, such as sudden emergence, complete disappearance, or kinematic changes of BAL troughs, are of particular interest, as they offer valuable clues about the complex nature of quasar winds \citep{vilko2001, Filiz2013,mcgraw2017, vivek2018, cicco2018, rogerson2018, lu2025}. Such significant changes, often observed over rest-frame timescales of months to years \citep[e.g.,][]{Filiz2013, Vivek2014}, are typically attributed to fluctuations in the quasar's ionizing flux and/or variations in the outflow's covering fraction, or transverse bulk motion within the outflow. 
In addition, it was also shown that changes in shielding gas column density originating from highly ionized outflows close to the accretion disk can lead to ionization changes in outflows located further away as demonstrated in the case of NGC 5548 \citep{kaastra2014, arav2015}. This mechanism becomes important when one sees highly correlated variability over the full absorption profile when there is  no appreciable change in the optical continuum flux.
Hence, it is important to follow-up BAL sources that show extreme BAL variability over rest-frame timescales of a few months to give us an opportunity to constrain the nature of these outflows \citep{vivek2018, Aromal2021} that are crucial in AGN feedback-galaxy evolution models.

In the literature, many studies have found interesting LoBALs which have shown appearance or disappearance signatures \citep{vivek2012,filiz2014, yi2019b, yi2021} and in some cases, even kinematic shifts \citep{lu2020}.
However, it is important to note that this extreme variability is very rare in LoBALs compared to their HiBAL counterparts \citep{yi2021}.
Also, it was shown that LoBALs tend to be seen at lower velocities relative to corresponding HiBALs like \civ\ that reach very high velocities of a few 10,000 \kms\ \citep{filiz2014}.
\citet{hamann2019} concluded that LoBALs form in harsh radiative environments, characterized by high ionization parameters (U) similar to HiBALs, but with significantly larger total column densities that enhance radiative shielding.
Using the spectral synthesis modeling software SimBAL, it has been shown that several LoBALs, particularly FeLoBALs, possess large total column densities (log($N_H$) up to 23), and span a wide range of ionization parameters and gas densities \citep{karen2018, karen2022}.
Interestingly, from an evolutionary picture, some studies interpret LoBALs as part of a short-lived stage in quasar evolution where the SMBH is evacuating large amounts of gas from its circumnuclear environment, otherwise known as the ``blowout" phase \citep{urrutia2009}. 

We have been undertaking a dedicated spectroscopic monitoring of 64 HiBALs showing high velocity (ejection velocities in excess of 15000 kms$^{-1}$) outflows using Southern African Large Telescope \citep[SALT;][]{aromal2023,aromal2024}. This combined with the SDSS spectra allows us to probe time variability over few months to $\sim$ 8 years in the quasar's rest frame.  
Here we study one of the interesting BAL quasar  (J115636.82+085628.9 hereafter J1156+0856, $z_{\mathrm{em}} = 2.1077$) in our sample that shows the maximum \civ\ BAL equivalent width variability that coincides with an episode of strong photometric light curve variations. 
The quasar has a bolometric luminosity of log$(L_{bol} \text{ [erg s}^{-1}]) = 46.75$ and a black hole mass of log$(M_{BH} / \msun) = 8.92$, both of which are within the typical range of quasar properties in our sample \citep{aromal2023}.
This source is also unique in that we observe both the emergence of LoBALs in \aliii\ and \mgii, and their subsequent disappearance during our monitoring (effectively a HiBAL-LoBAL-HiBAL transition), which is an extremely rare occurrence.
This LoBAL absorption emerges at high velocities of 20,000 \kms\ ($\sim$ 0.07 c) and such cases have not been previously reported in the literature.
We find that the LoBAL variability is closely associated with HiBAL variations as seen in \civ\ and \siiv\ BALs.
Such variability is rare in nature and offers us an unique opportunity to study the origin of LoBALs and their connection to their HiBAL counterparts.
It will also help us in understanding what triggers BAL variations as these different ionized species may come from different parts of the outflow which may respond differently to any global mechanism that leads to these variations.

The paper is organized as follows.
In Section~\ref{sec:obs}, we present the observations used in our study.
Section~\ref{sec:analysis} provides details of results based on the absorption line variability in \J11.
Section~\ref{sec:discussions} presents the discussions and 
in Section~\ref{sec:conclusions}, we discuss the main results and provide a summary of our work.
Throughout this paper we use the flat $\Lambda$CDM cosmology with  $H_0$ = 70 \kms\ Mpc$^{-1}$ and $\Omega_{m,0}$ = 0.3.

\section{Observations $\&$  Data used in this study}
\label{sec:obs}

We obtained two high-quality, fully reduced spectra of \J11\ from SDSS/BOSS (Sloan Digital Sky Survey/ Baryon Oscillation Spectroscopic Survey) archives, data release 16 \citep{lyke2020}. The SDSS (BOSS) spectra covers a wavelength range of 3800-9200 (3600-10400) \AA\ at a resolution, R $\sim$ 2000 (150 \kms).
Using the Southern African Large Telescope (SALT), we carried out spectral monitoring of \J11\ from 2022-2024 and obtained three spectra with almost an year separation in the observed frame.
For this, we used the Robert Stobie Spectrograph \citep[RSS,][]{Burgh2003, Kobulnicky2003}
at SALT in long-slit mode using a 1.5" wide slit and the PG0900 grating.
For grating angle = 14.375, this setting provides a spectrum with wavelength coverage of 3900 \AA\ to 7000 \AA\ excluding the two CCD gap regions at 4935-5000 \AA\ and 5980-6050 \AA.
Details of all available spectra are given in Table~\ref{tab_obs}. The shortest and longest time-scale probed at the quasar's rest frame are 0.28 and 6.7 yrs respectively.

Continuum fitting of the observed spectra is very important in BAL variability studies. 
We used the publicly available multi-component spectral fitting code {\sc PyQSOFit}\footnote{https://github.com/legolason/PyQSOFit} \citep{guo2018a} 
to fit the continuum and broad emission lines (BEL) in our spectra.
We identified wavelength ranges devoid of any absorption lines and then fitted these regions with a power-law + multiple Gaussian (for BELs) model which provided fairly good fits using $\chi^2$ minimization as shown in Fig~\ref{fig:J1156_subplots}.

We also characterize the continuum variability of the source using broad-band photometric light curves.
For this, we obtained publicly available light-curves of \J11 from the Panoramic Survey Telescope and Rapid Response System \citep[Pan-STARRS;][]{panstarrs2016}, the Palomar Transient Factory \citep[PTF;][]{Law2009} and the Zwicky Transient Facility \citep[ZTF;][]{zptf2019b,zptf2019a} surveys. Pan-STARRS data includes five broad band filters, i.e., g, r, i, z and y whereas ZTF uses the g, r and i bands. Details of photometric variability of \J11 are discussed in sub-section~\ref{subsec:light_curve}.

\begin{deluxetable}{ccccccccc}
\tablecaption{Log of observations and details of spectra obtained at different epochs \label{tab_obs}}
\tablecolumns{8}
\tablewidth{0pt}
\tablehead{
\colhead{Epoch} &
\colhead{Telescope} &
\multicolumn{2}{c}{Date of observations} &
\colhead{$\Delta t_{\text{rest}}$\tablenotemark{a}} &
\colhead{Exposure } &
\colhead{Spectral res.} &
\colhead{Wavelength} &
\colhead{S/N\tablenotemark{b}} \\
\colhead{} & \colhead{} &
\colhead{(M/D/Y)} & \colhead{(MJD)} &
\colhead{(yr)} & \colhead{time (s)} & \colhead{ (\kms)} & \colhead{range (\AA)} & \colhead{}
}
\startdata
1 & SDSS & 04/04/2003 & 52733 & - & 2340 & 150 & 3808$-$9212  & 11.32 \\
2 & SDSS & 02/03/2012 & 55960 & 2.84 & 3603 & 150 & 3555$-$10344 & 17.79 \\
3 & SALT & 04/21/2022 & 59690 & 6.13 & 2300 & 304 & 3897$-$6980  & 17.64 \\
4 & SALT & 02/27/2023 & 60002 & 6.40 & 2400 & 304 & 3897$-$6980  & 26.45 \\
5 & SALT & 02/06/2024 & 60346 & 6.71 & 2400 & 304 & 3907$-$6984  & 26.14 \\
\enddata
\tablenotetext{a}{Time difference between spectroscopic epochs in rest-frame with respect to the first epoch.}
\tablenotetext{b}{Signal-to-noise ratio per pixel calculated over the wavelength range 5000--5200 \AA.}
\end{deluxetable}

\section{Absorption line variability}
\label{sec:analysis}

\begin{figure*}
    \centering
    \includegraphics[viewport=90 40 1200 1250, width=0.85\textwidth,clip=true]{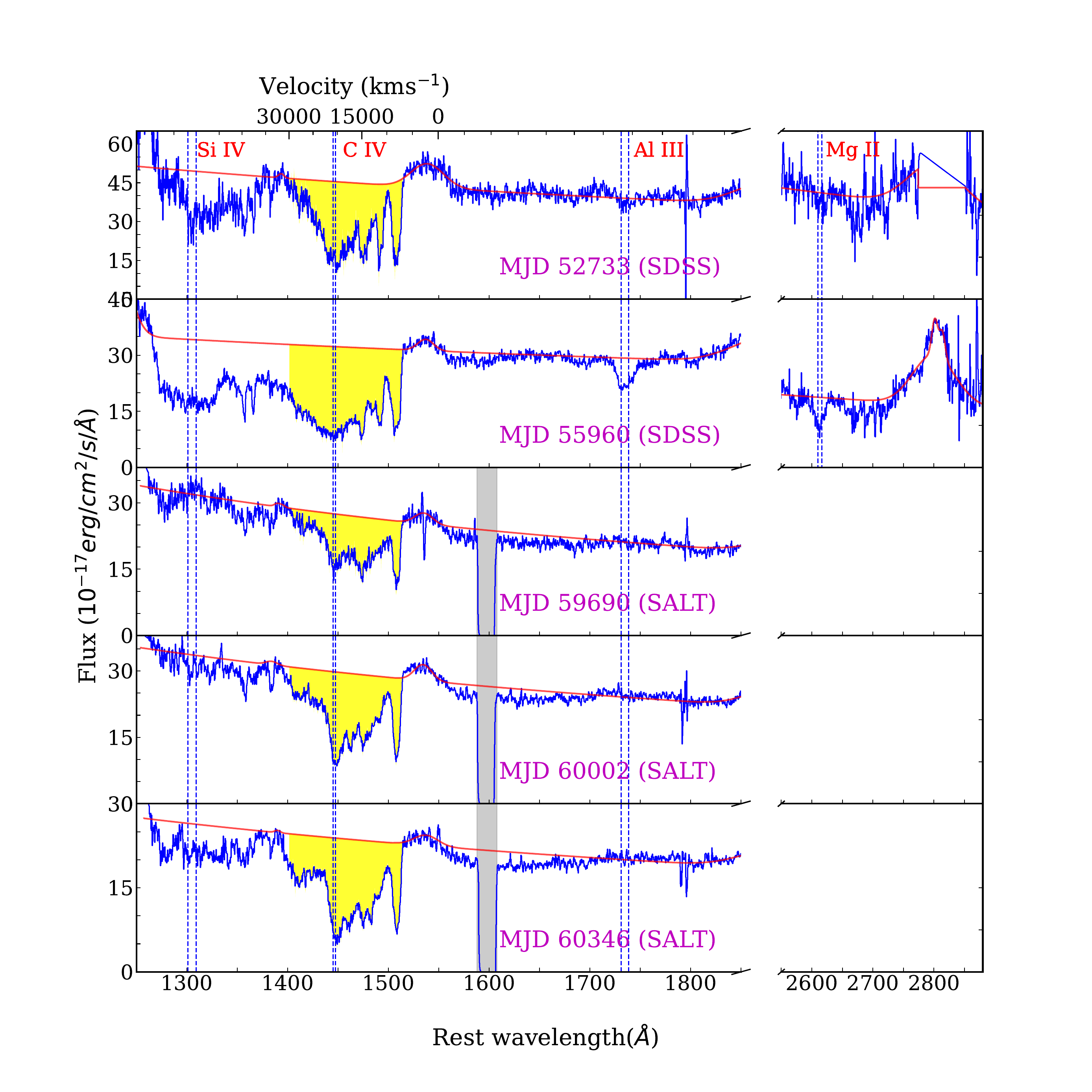}
    \caption{
    The rest-frame spectra of \J11\ (blue) along with the best continuum fit (red) defined with respect to \zem = 2.1077 are shown.
    Absorption line variability of \civ\ broad components (up to v = 30000 \kms) identified with yellow-shaded regions is apparent. 
    The velocity scale for \civ\ BAL absorption with respect to the systemic redshift (\zem = 2.1077) is provided at the top. 
    The blue dashed lines represent the absorption in different ionized species for BAL component `D' (v$\sim20000$ \kms) where LoBAL absorption emerges during epoch-2 (refer subsection~\ref{subsec:bal_regions}).
    Gray-shaded regions represent the CCD gaps in SALT spectra. 
    }
    \label{fig:J1156_subplots}
\end{figure*}

Figure~\ref{fig:J1156_subplots} shows the spectra of \J11\ obtained during five different epochs (shown in blue lines) together with the best-fit continuum (shown in red lines). 
In addition, the continuum normalised spectra covering the \civ\ and \siiv\ absorption are compared in Figure~\ref{fig:J1156_norm_overplots}.
The details of \civ\ absorption line based measurements are provided in Table~\ref{tab_ew}.
As mentioned above, the
spectra from the initial two epochs (shown in the top two panels) are from SDSS and the spectra for the subsequent three epochs were obtained using SALT.
The \civ\ broad absorption line (yellow shaded region in each panel) in \J11\ span a velocity range that starts at v$_{min}$ $\sim$ 6700 \kms\ and extends up to the conservative upper limit of 30000 \kms. This is according to the definition given in \citet{Weymann1991} where a BAL is defined as a continuous absorption wider than 2000 \kms\ below 90 $\%$ of the continuum level.
There is a strong possibility that absorption is present at even higher velocities than this limit (especially during the second epoch), but for the ease of discussions and to avoid possible overlap with \siiv\ absorption, we set the maximum BAL velocity v$_{max}$ to 30000 \kms\ in our analysis. 
The main trends we observe from Figure~\ref{fig:J1156_norm_overplots} are,
\begin{enumerate}
    \item {}  The \civ\ BAL region is extremely variable and undergoes non-uniform depth variations between any pair of spectroscopic epochs considered. This is confirmed by the rest equivalent width variations from column 2 in Table~\ref{tab_ew}.
    \item{} The \civ\ absorption is strongest (i.e with a rest equivalent width of $65.16\pm0.29$ \AA) during epoch-2 (i.e MJD 55960). During this epoch we also see strong, broad, and saturated \siiv\ absorption. A narrow Al~{\sc iii} and Mg~{\sc ii} absorption with the ejection velocity of $18000-23000$ \kms\ and $19000-22500$ \kms\ respectively that coincides with the velocity range where deepest \civ\ absorption is present. This component is identified with the vertical blue dashed lines in Fig~\ref{fig:J1156_subplots}.
    
    \item{} The \siiv\ absorption observed during epoch 2 is broad with a flat-bottomed profile, indicative of line saturation and a covering factor of $f_c \sim 0.5$. Here, we assume that the absorption depth of the saturated \siiv\ BAL in the continuum normalized spectrum corresponds to its covering fraction. However, the \civ\ absorption depth is higher than that of \siiv\ suggesting covering factor of \siiv\ being smaller than that of \civ. 

    \item{} The \civ\ absorption is at its weakest strength (i.e rest equivalent width of 30.43$\pm$0.35\AA) during epoch-3 (i.e MJD 59690) and \siiv\ and \aliii\  absorption lines are not detected in this epoch. Unfortunately, the wavelength range of \mgii\ absorption is not covered by our SALT observations. It is also evident from Fig.~\ref{fig:J1156_norm_overplots} that some of the narrow \civ\ absorption components clearly visible during epoch-1 and epoch-2 are not visible in the spectrum taken during epoch-3.
    
    \item{} The \civ\ absorption shows an increasing trend in the following two epochs, i.e. epoch 4 and 5. The overall profile looks roughly similar
    (compared to epoch-1) except for a distinct component emerging around an ejection velocity of 25000-30000 \kms. However, some of the narrow components seen in the first two epochs do not reappear during this period.
\end{enumerate}

Thus, the variability of the \civ\ absorption line shows interesting trends. To gain further insight, we next study the variability of the absorption strength as a function of the outflow velocity. For this, we adopt a method where we divide the entire BAL profile into a few specific regions of interest based on the overall variability and study them in detail in the following subsections.

\subsection{BAL regions}
\label{subsec:bal_regions}
\begin{figure*}
    \centering
    \includegraphics[viewport=120 20 1580 955, width=0.85\textwidth,clip=true]{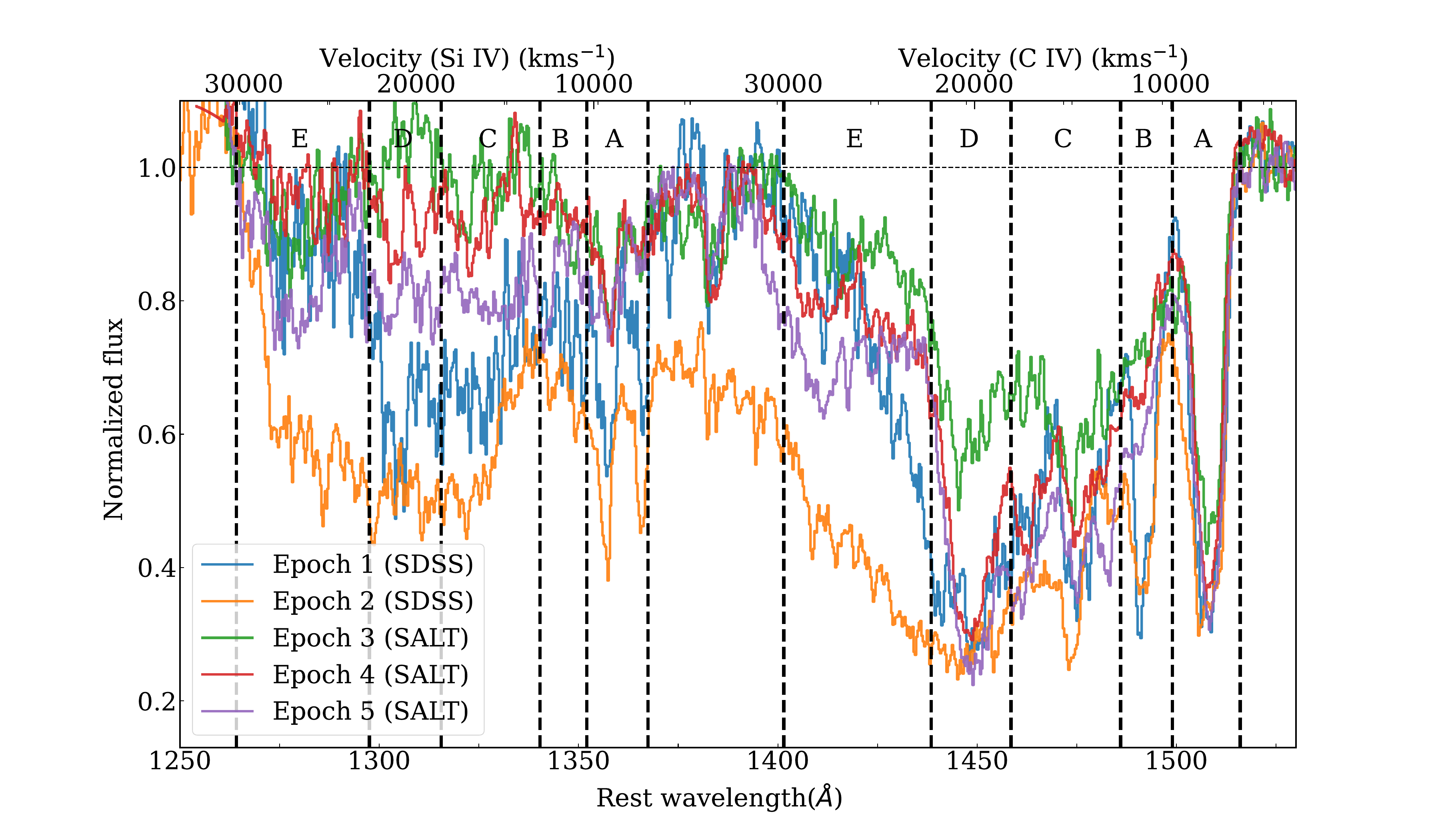}
    \caption{The figure shows the overlaid normalized spectra of \J11\ in the five epochs focusing on regions blueward of \civ\ BEL that covers \civ\ and \siiv\ BAL regions. The black dashed lines indicate the distinct BAL regions for \siiv\ (left) and \civ\ (right), as described in Section~\ref{subsec:bal_regions}.}
    \label{fig:J1156_norm_overplots}
\end{figure*}

We split the entire \civ\ BAL profile, as observed in epoch 1,
into five regions after carefully considering the diverse nature of profile variability 
during our monitoring.
The velocity ranges for regions A, B, C, D, and E are 6,700--10,000 \kms, 10,000--12,500 \kms, 12,500--18,000 \kms, 18,000--22,000 \kms and 22,000--30,000 \kms, respectively, as shown for both the \civ\ and \siiv\ absorption  in Fig~\ref{fig:J1156_norm_overplots}.
Again, these regions are chosen keeping in mind the interesting variability shown at subsequent epochs.

Now, let us look at the BAL profile observed in the first epoch (see Fig~\ref{fig:J1156_norm_overplots}).
Both regions A and B consist of a strong, narrow component with a distinct absorption profile in the first epoch. 
We detect corresponding \siiv\ doublets for region A whereas region B lacks the same despite having a comparable \civ\ absorption depth. This may indicate the overall column density of gas contributing to A being much larger than that of B. The \civ\ absorption strength being equal
can be related to line saturation combined with partial coverage.
In region C, multiple \civ\ narrow components are visible in the first epoch and they show extreme variability in both their position and profile shape in subsequent epochs, as discussed later in detail.
Region D consists of the strongest distinct \civ\ absorption with a possible unresolved \siiv\ absorption present at the corresponding velocities.
Region E, which is at the highest velocities, does not have any distinct absorption components, but a smooth absorption is present throughout the region which gradually decreases in strength towards high-velocity end. 
Now, let us look at the variability in each defined BAL region and study in detail the nature of the BAL outflow in \J11.

\begin{figure}
    \centering
    \includegraphics[viewport=55 85 2420 1550, width=\textwidth,clip=true]{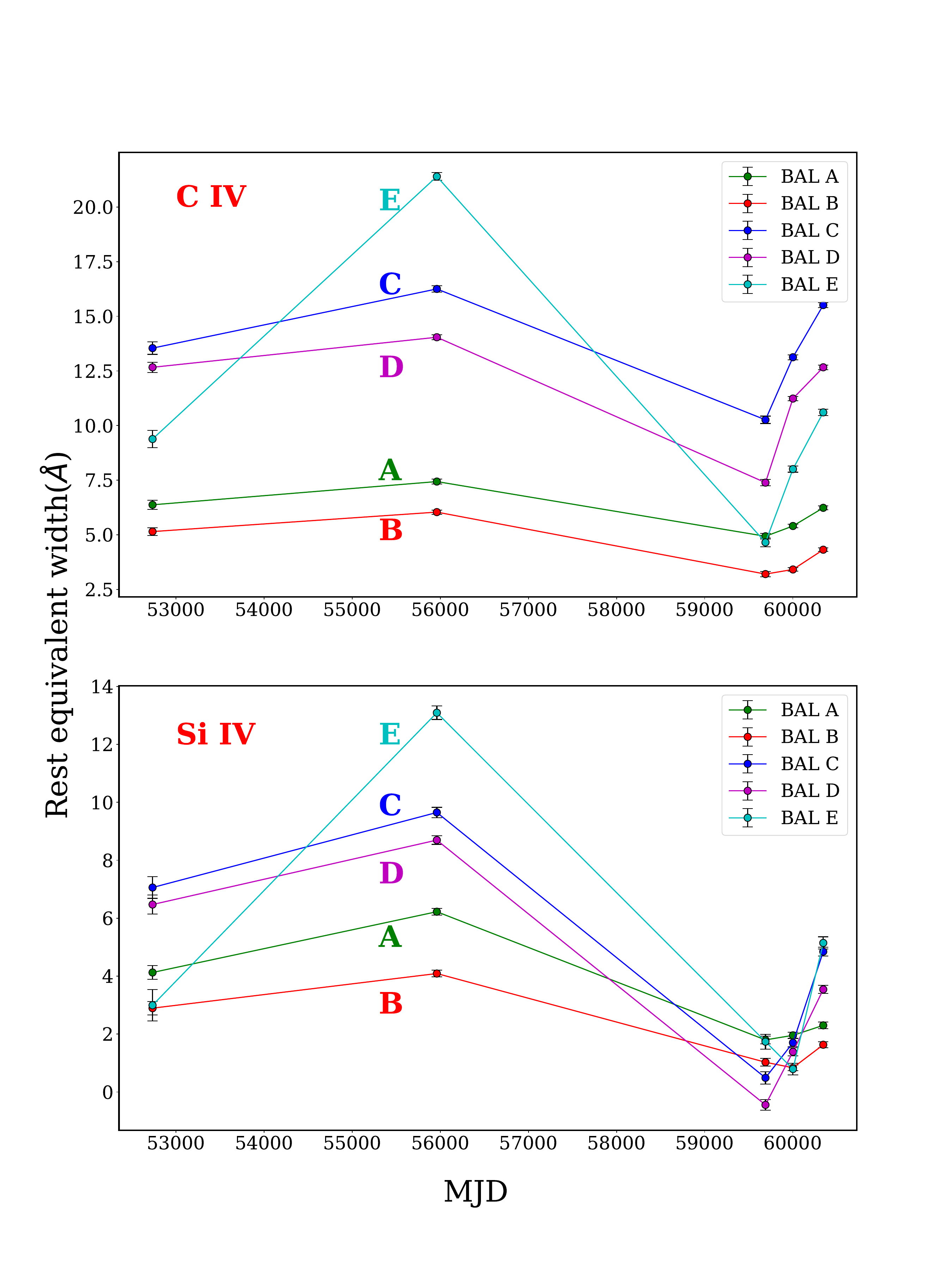}
    \caption{This figure shows the \civ\ and \siiv\ equivalent widths calculated for different BAL regions as defined in Section~\ref{subsec:bal_regions} as a function of time.
    }
    \label{fig:eqw_plot}
\end{figure}

\begin{deluxetable*}{cccccccc}
\tablecaption{Log of equivalent widths calculated at different epochs \label{tab_ew}}
\tablecolumns{8}
\tablewidth{0pt}
\tablehead{
\colhead{Epoch} &
\colhead{$W_{\text{\civ}}$} &
\colhead{$\frac{dW_{\text{\civ}}}{dt}$\tablenotemark{a}} &
\colhead{$W_{\text{\civ}}^A$\tablenotemark{b}} &
\colhead{$W_{\text{\civ}}^B$\tablenotemark{c}} &
\colhead{$W_{\text{\civ}}^C$\tablenotemark{d}} &
\colhead{$W_{\text{\civ}}^D$\tablenotemark{e}} &
\colhead{$W_{\text{\civ}}^E$\tablenotemark{f}} \\
\colhead{} &
\colhead{(\AA)} &
\colhead{(\AA/yr)} &
\colhead{(\AA)} &
\colhead{(\AA)} &
\colhead{(\AA)} &
\colhead{(\AA)} &
\colhead{(\AA)}
}
\startdata
1 & 47.11 $\pm$ 0.60 & ---     & 6.37 $\pm$ 0.20 & 5.14 $\pm$ 0.18 & 13.54 $\pm$ 0.28 & 12.66 $\pm$ 0.23 & 9.38 $\pm$ 0.39 \\
2 & 65.16 $\pm$ 0.29 & 6.34    & 7.43 $\pm$ 0.11 & 6.03 $\pm$ 0.09 & 16.25 $\pm$ 0.14 & 14.04 $\pm$ 0.11 & 21.39 $\pm$ 0.18 \\
3 & 30.43 $\pm$ 0.35 & --10.56 & 4.93 $\pm$ 0.12 & 3.20 $\pm$ 0.12 & 10.26 $\pm$ 0.17 & 7.38  $\pm$ 0.15 & 4.65  $\pm$ 0.20 \\
4 & 41.18 $\pm$ 0.23 & 39.07   & 5.40 $\pm$ 0.08 & 3.41 $\pm$ 0.08 & 13.12 $\pm$ 0.11 & 11.23 $\pm$ 0.09 & 8.00  $\pm$ 0.14 \\
5 & 49.33 $\pm$ 0.25 & 26.87   & 6.23 $\pm$ 0.09 & 4.32 $\pm$ 0.08 & 15.51 $\pm$ 0.11 & 12.66 $\pm$ 0.10 & 10.60 $\pm$ 0.15 \\
\enddata
\tablenotetext{a}{  $\frac{dW_{\text{\civ}}}{dt}$ is the rate of change of equivalent width of the entire BAL region (including A, B, C, D and E components) with respect to the previous epoch in rest-frame time scale.}
\tablenotetext{b,c,d,e,f}{\hspace{1.2cm}$W_{\text{\civ}}^A$, $W_{\text{\civ}}^B$, $W_{\text{\civ}}^C$, $W_{\text{\civ}}^D$ and $W_{\text{\civ}}^E$ are the \civ\ rest equivalent width of region A, B, C, D and E respectively, integrated over a distinct portion of the absorption profile.}
\end{deluxetable*}

\subsection{Equivalent width variations}

In terms of the variations of the equivalent width (W), \J11 shows one of the most extreme changes in absorption strength ever reported in BAL quasars.
Interestingly, these variations are not uniform across the BAL profile, and analyzing variability in different BAL regions helps reveal a clear picture of its unique nature.
Figure~\ref{fig:J1156_norm_overplots} shows the interesting depth variations in each BAL region and Figure~\ref{fig:eqw_plot} and Table~\ref{tab_ew} show the equivalent widths calculated for each BAL region (as defined in Section~\ref{subsec:bal_regions}) at different epochs. Now, we analyze the BAL variability in detail in the following paragraphs.

We described the BAL profile in epoch 1 in the previous sub-section~\ref{subsec:bal_regions}.
In the second SDSS epoch ($\sim$ 3 yr in rest-frame after epoch 1), we see a large positive change in the equivalent width ($\Delta$W $\sim$ 18 \AA) especially at velocities greater than 20,000 \kms. 
In this epoch, regions A and B show little variation in \civ\ although corresponding \siiv\ absorption has increased considerably. However, caution should be exercised for these regions since possible \civ\ absorption exceeding 30,000 \kms\ can overlap with \siiv\ absorption and lead to higher values of \siiv\ equivalent width. 
Region C shows a large variation with the emergence of a narrow component at 13000 \kms\ whereas an already existing absorption component at 15000 \kms\ becomes narrower and deeper.
The regions D and E show significant depth variations where E shows an increase in depth as large as $\sim$0.5 at $\sim$25000 \kms.
Interestingly, for region D, in addition to \siiv\ absorption becoming deeper, a considerable \aliii\ and \mgii\ absorption also emerges at $\sim$ 18000 - 22000 \kms.
The equivalent widths of the new \aliii\ and \mgii\ absorption are 4.77 $\pm$ 0.16 \AA\ and 2.63 $\pm$ 0.24 \AA\ respectively.
It is interesting to note that the low-ionization absorption emerges in a narrow region at relatively large velocities. This is opposite to what we expect based on studies such as \citet{filiz2014} which found that the \aliii\ BALs generally correspond to the low-velocity portions of the \civ\ trough.
Also, there is little difference in the peak absorption depth (at v $\sim$ 20000 \kms) in region D between epochs 1 and 2, despite the emergence of LoBALs at corresponding velocities, indicating the strong saturation in \civ\ absorption. 
The region E shows the strongest variations ($\Delta W$ $\sim$ 12 \AA) with significant changes in depth accompanied by a strong \siiv\ absorption appearing at these velocities for the first time during our monitoring.
These variations point towards a possibly large increase in the line-of-sight column density across the entire profile during epoch 2.
This can be associated with the eruption of a strong high-velocity wind from the accretion disk as implied by the large velocity gradient of this new absorption or the transit of a dense flow component into our line of sight at relatively smaller distances, as discussed later in Section~\ref{sec:discussions}.

Epoch-3 using SALT is observed after another $\sim$3 yr in rest-frame since epoch-2. Once again, we observe dramatic changes in the absorption, but this time the entire BAL weakens significantly ($\Delta$W $\sim -35$ \AA) and reaches the lowest equivalent width observed. 
Region A shows the least depth variations confirming its highly saturated nature. But the resolved \siiv\ doublet from region A weakens and also reaches a doublet ratio close to 2:1 indicating the \siiv\ absorption may have become optically thin in this epoch.
More interestingly, the distinct \civ\ narrow component in region B completely disappears.
Both C and D show a large decrease in \civ\ and \siiv\ depth with distinct narrow components in these regions almost disappear.
It is interesting to note that this also leads to the complete disappearance of \aliii\ absorption at $\sim$ 20000 \kms.  We are unable to comment on the further evolution of \mgii\ absorption in this study as the SALT spectra do not cover this wavelength region.
Region E also undergoes a significant weakening and reaches depths ($\sim$ 0.15) smaller than those of epoch 1. 
Overall these variations contributes to one of the highest changes in \civ\ BAL equivalent width ($\sim$ 35 \AA) reported in the literature \citep[see figure 13 of][where the highest absolute equivalent width changes are $\sim$ 15\AA\ in one of the largest BAL quasar samples.]{Filiz2013}.

In epoch-4 followed by epoch-5, the BAL again goes through a strengthening episode ($\Delta$W $\sim 8$ \AA) albeit not as dramatic as the one observed in epoch-2. 
It is interesting to note that this change occurs on relatively short time scales of a few months in the rest-frame, thanks to our close monitoring using SALT.
The BAL equivalent width increases considerably with one of the largest rates of change of equivalent widths ($\sim$ 39 \AA$/$yr between epoch-3 and -4 in rest-frame time scales, see Table~\ref{tab_ew} and Fig~\ref{fig:eqw_plot}).
In these later epochs, the absorption profile roughly becomes similar to that in epoch-1 except for 
(1) the \civ\ narrow component in region B does not re-emerge despite small changes in the overall average depth,
(2) the narrow components in region C undergo significant variability with possible kinematic shift signatures 
(3) the width of the broad component in region D decreases considerably while retaining roughly the same absorption depth at v $\sim$ 20000 \kms as in epoch-1. However, it is interesting to note that the \aliii\ absorption present in epoch-2 never reappears and
(4) In region E, a new absorption component emerges at 25000-30000 \kms\ in epoch 4 which subsequently becomes stronger in epoch 5.
The restoration of the overall absorption profile after a strengthening and weakening episode suggests that an absorber component may be always present along our line-of-sight possibly varying in its ionization structure.
This suggests that the relative distribution of \civ\ absorber column density as a function of velocity remains roughly similar in the first and last epochs, despite the appearance of strong absorption between epochs 1 and 2. 
That said, the relatively minor changes in the absorption profile between these epochs are likely due to residual wind components that persist even after the disappearance of a large-scale outflow component from our line of sight.
We discuss the possible reasons for these variations in Section~\ref{sec:discussions}.

\subsection{Light curve variations and their connection to BAL variability}
\label{subsec:light_curve}

\begin{figure}
    \centering
    \includegraphics[viewport=30 10 2070 680, width=\textwidth,clip=true]{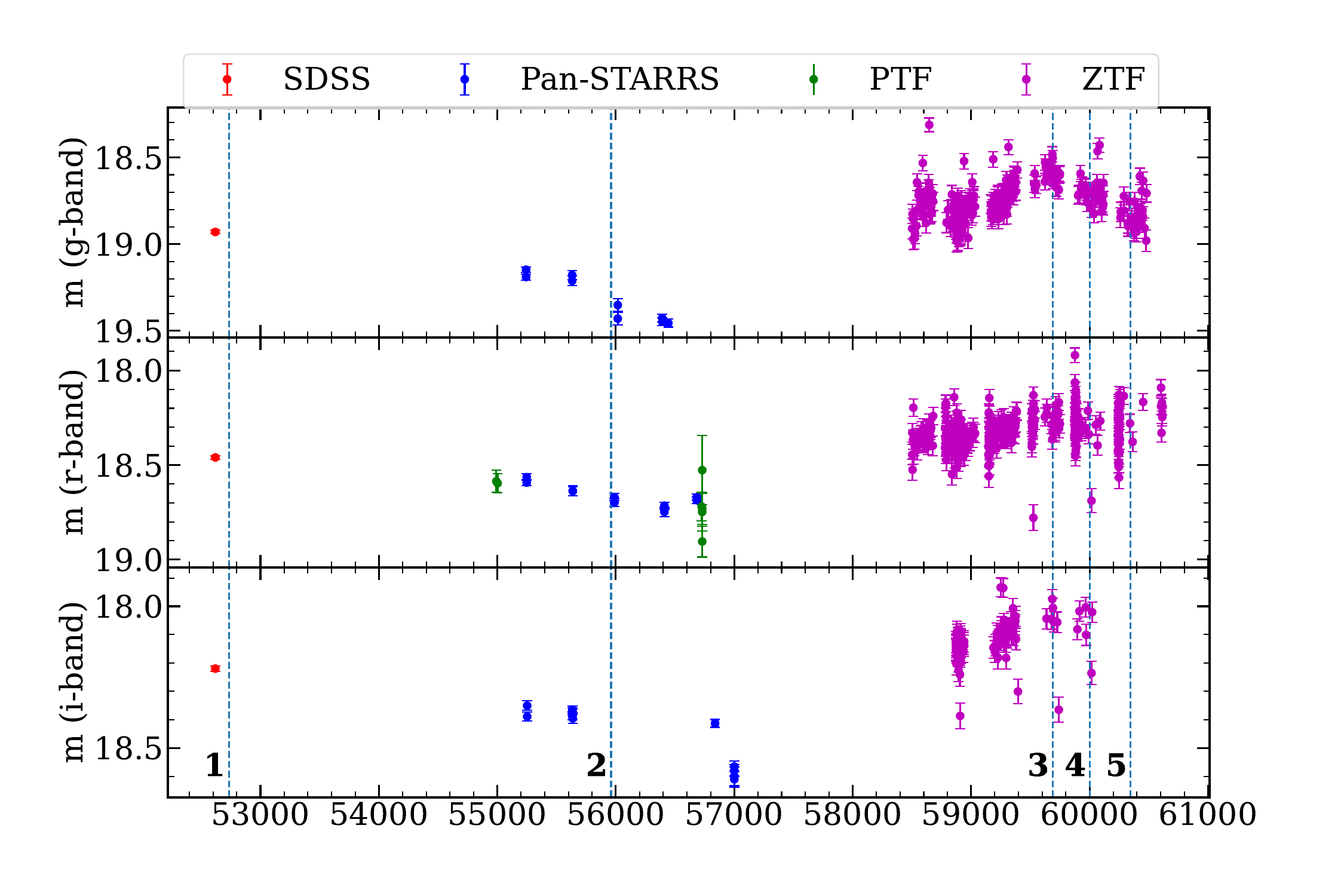}
    \caption{The light curves of \J11\ in g (top), r (middle) and i (bottom) optical bands are shown as obtained from publicly available photometric surveys. 
    The spectroscopic epochs are marked (vertical blue dashed lines) with the epoch numbers marked in the bottom panel.
    }
    \label{fig:lc}
\end{figure}

In Fig.~\ref{fig:lc}, we show the available light curves of \J11\
obtained in three different filters (g, r, and i-bands).  
These light curves are not sampled uniformly in time,  but the ZTF measurements provide good coverage during our last three spectroscopic epochs.
We find significant variations in the photometric magnitudes in all bands between our spectroscopic epochs as shown in Fig.~\ref{fig:lc}.
We ignore the relatively minor differences in filter transmission functions between surveys when comparing magnitudes in our analysis. 
This approach is justified by findings that the mean offsets in $g$- and $r$-band magnitudes between SDSS, Pan-STARRS, and ZTF are $\sim$0.05 mag at the redshift of \J11\ \citep[see Figure 1,][]{wu2024}, which is negligible compared to the large magnitude variations observed in our source and does not impact any of our conclusions.

In epoch-2, as stronger absorption emerges, the source becomes dimmer by $\sim$0.5 ($\sim$0.3) magnitude in g- (r-) band compared to epoch-1. 
Subsequently, the ZTF points close to epoch 3 show a significant brightening of the order of $\sim$ 1 mag and $\sim$ 0.5 mag in the g- and r-bands respectively as the BAL weakens.
Although these changes are consistent with the BAL variability (in the g-band), the decrease in depth of the BAL alone cannot account for such large variations in magnitudes observed.
After epoch 3, we notice a gradual dimming in the g-band (the same is not evident in r-band) until epoch-4.
These trends indicate
that the absorbing gas maybe responding to the changes in continuum flux and an increase in continuum leads to the weakening of BAL features. 
This anti-correlation trend has been extensively reported in the literature 
\citep{wang2015, vivek2019, mishra2021}. 
This would suggest that in the case of \J11, photoionization induced variations are important to explain the BAL variability.
However, it is good to keep in mind that the changes in UV flux (as probed by g-band) and high energy X-ray radiation, which can affect the ionization levels in the cloud, need not be correlated with each other \citep[see for example,][]{arav2015}. This prevents us from drawing any serious conclusion with just UV/optical light curves in the absence of X-ray observations.
In the next section, we model the variations combining spectral and photometric variations with photo-ionization simulations.

\section{Discussions}
\label{sec:discussions}

\subsection{The appearance of a new outflow component?}
\label{subsec:discussion1}

One of the primary drivers of BAL variability is the changse in the ionization structure of the outflow, induced by fluctuations in the ionizing flux incident on the absorber \citep{grier2015ApJ, aromal2022}. Alternatively, variability can result from changes in the covering fraction as the outflow transits our line of sight (LOS) or due to its bulk motion away from the accretion disk \citep{Capellupo2013iii, mcgraw2015, green2023}. Previous studies suggest that the underlying causes are often more complex, with both ionization changes and geometric effects contributing to the observed variability in many quasars \citep{mcgraw2017, Aromal2021}.

Another plausible scenario, which is less explored, involves the presence of multiple outflow components, either located at varying distances from the disk or as distinct regions within the same outflow, separated on larger spatial scales. 
In such cases, the observed variability may depend on interactions between these different flow components \citep[see for example,][]{arav2015}. 
In this study, we explore this multicomponent outflow scenario for \J11 and argue that it provides a more comprehensive explanation for the observed variability than a simple photoionization-driven model involving a single outflowing cloud. 
While our analysis does not rule out the alternative mechanisms, we propose that this new picture—motivated by several intriguing observational features of \J11—provides an equally viable explanation that aligns well with previous findings.

First, we consider either the emergence of a new outflow from the accretion disk or the transit of a strong outflow component into our line of sight at a relatively smaller distance, that can lead to several interesting consequences for variability in a BAL outflow that is already present.
The newly entered flow material can produce \civ\ absorption in favorable conditions
and also significantly affect the ionization structure of a previously existing outflow at larger distances due to efficient shielding \citep[see for example ][]{fukumura2024ApJ}.
Effective shielding of ionizing radiation by this gas  can prevent high-energy photons from reaching distant clouds which can lead to an increase in the fraction of low-ionized species such as \aliii\ and \mgii.
Here, the variability will be difficult to explain as the two effects, i.e. the addition of a new absorbing material along our line of sight and the possible changes in the ionization structure of already existing outflows, play an important role in determining changes in the optical depth of various ionized species.
Previous studies such as \citet{vivek2018} have shown that the column density ratios in BAL absorbers are highly sensitive to the shielding gas column density using {\sc Cloudy} simulations.
In addition, \citet{hamann2019} have clearly shown that LoBAL absorption arises from a highly shielded region in high column density outflows while comparing Hi-BAL and LoBAL quasars with \pv\ absorption.
Here, we develop a model that tests the plausibility of a shielding scenario that may be responsible for the emergence of LoBALs as observed in \J11.
In this scenario, the shielding gas absorbs a substantial fraction of the radiation from the accretion disk, thereby changing the photoionization conditions of the gas located farther out. We propose that this shielding can be caused by an additional flow that transits our line of sight during epoch 2 as explained in detail below. 
Note that this may be slightly different from a conventional definition of shielding gas used in the literature before, since we assume that this new flow can also lead to absorption in UV lines.

For our source, we assume that the absorption present in epoch-1 is due to an outflow that is already present along our LOS, say, flow-I.
Now, we can also assume that flow-I may not be a single structure, but consists of kinematically separate sub-components corresponding to the different BAL regions as discussed in subsection~\ref{subsec:bal_regions}.
During epoch 2, a new flow having large column density either emerges near the accretion disk or transits into our line of sight at a relatively small distance, referred to hereafter as flow-II. 
In the following discussion, we use flow-II to represent both scenarios, as their impact on the observed variability, primarily through providing shielding for flow-I, is effectively indistinguishable for the purposes of our analysis.

\begin{figure}
    \centering
    \includegraphics[viewport=15 10 1500 450, width=\textwidth,clip=true]{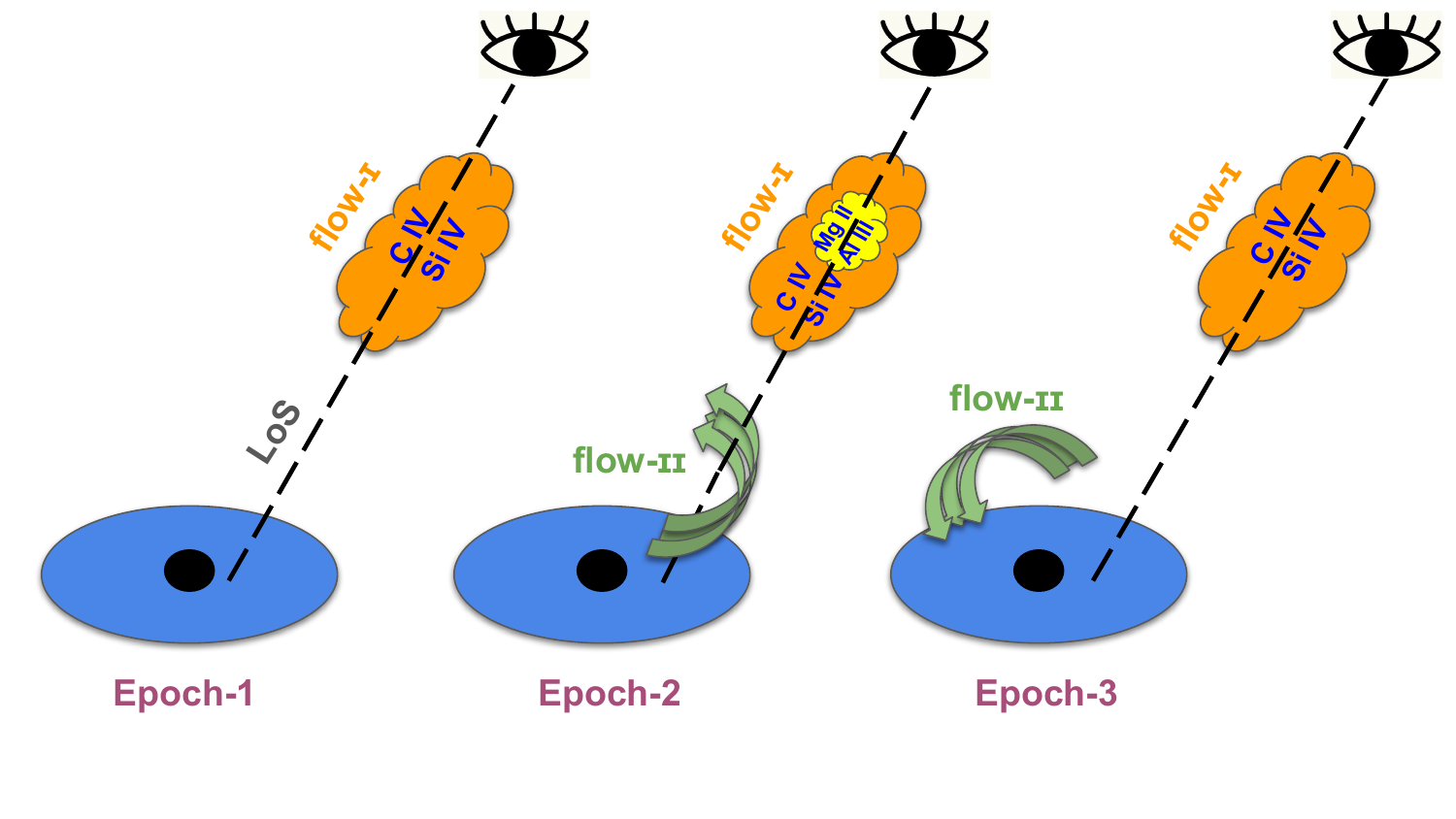}
    \caption{This figure shows a simple schematic diagram of the model of an accretion disk (blue) + two flow components, namely flow-I and flow-II that we propose in order to explain the BAL variability in \J11. The three panels represent the first three epochs where flow-I is present in the first epoch, then flow-II enters our LoS in epoch-2 which leads to LoBAL absorption in flow-I and later moves out of the LoS by epoch-3.
    }
    \label{fig:cartoon}
\end{figure}

One of the primary motivations for invoking a multi-component outflow scenario arises from the distinct absorption properties observed in region E. 
Nearly all of the \siiv\ absorption in this region appears during epoch 2, and this can be attributed to the emergence of a new component, i.e., flow-II. 
Assuming that the roughly flat-bottomed \siiv\ absorption profile in epoch 2 is a result of saturation, we estimate a covering fraction of $C_{v, \text{\siiv}}(\text{flow-II}) \sim 0.4$, corresponding to the average absorption depth.
By contrast, the \civ\ absorption in region E exhibits a sharp decline in depth, from approximately 0.7 to 0.4, toward higher velocities. 
This suggests that the average absorption depth, and thus the minimum total covering fraction, is significantly greater for \civ\ than for \siiv. 
Furthermore, the average change in \civ\ absorption depth between epoch 2 (when flow-II is present) and epoch 3 (when it nearly disappears) is also $\sim$0.4, closely matching $C_{v, \text{\siiv}}(\text{flow-II})$.
These findings imply that flow-II is a highly saturated, partially covering absorber in both \civ\ and \siiv, with $C_v \sim 0.4$. The additional \civ\ absorption in region E can be attributed to the already existing flow-I, which likely has a higher covering fraction and is superimposed on the saturated profile from flow-II. 
The disappearance of \siiv\ absorption in region E by epoch 3 further suggests that flow-I is not contributing to \siiv, indicating either a lower column density or different ionization state for flow-I at the high velocities. 
This scenario naturally accounts for the observed discrepancies in covering fractions between \civ\ and \siiv, supporting the presence of multiple outflow components along the line of sight, each with distinct physical properties and temporal behavior.
A simple schematic diagram that explains our model is shown in Figure~\ref{fig:cartoon} for better clarity.

According to this scenario, the appearance of flow-II in epoch 2 leads to effective shielding of the more distant flow-I from the central ionizing source, resulting in the sudden appearance of low-ionization BALs (LoBALs), such as \aliii\ and \mgii, at velocities corresponding to the strongest \civ\ absorption. 
Due to its proximity to the accretion disk, flow-II may move out of our line of sight on relatively short timescales (see discussion on transit timescales at the end of this subsection), thereby exposing flow-I once again to unattenuated radiation from the disk. 
It is also plausible that flow-II may have swept away or disrupted the shielding material near the disk during its passage, further enhancing the exposure of flow-I to high-energy photons.
This re-exposure could lead to a reduced \civ\ ion fraction in flow-I, consistent with the weakened absorption observed in epoch 3. However, in epochs 4 and 5, the absorption strengthens and roughly recovers its original profile seen in epoch 1, suggesting that the ionizing flux reaching flow-I may have returned to its earlier state. 
Interestingly, this period also coincides with the appearance of a distinct, high-velocity absorption component at 26,000–30,000 \kms. If this new feature is associated with the emergence of another flow launched close to the disk or transits our LOS, it may again provide the shielding necessary for LoBAL formation, potentially leading to the reappearance of LoBAL features in the near future. To test this hypothesis, we plan to continue short-timescale monitoring of this source.

We argue that the proposed multi-component outflow and a shielding scenario provides a consistent explanation for the observed BAL variability, based on the following points:
\begin{itemize}

\item In region D, the \aliii\ absorption does not reappear after epoch 2 (where it initially emerged) even though the corresponding \civ\ absorption recovers to similar depths in epochs 4 and 5. This behavior is difficult to explain with models invoking ionization changes driven solely by variations in the continuum flux in a single-component outflow. Furthermore, LoBAL features are generally observed at lower velocities, where line-of-sight column densities are expected to be higher \citep{filiz2014}. Hence, their sudden emergence at high velocities is better explained by a shielding scenario.

\item Except for epoch 2, the relative absorption depths in the \civ\ BAL profile remain largely consistent across the other epochs, indicating that the overall absorption profile is nearly restored to its original shape seen in epoch 1. If the strong variability were primarily driven by large changes in column density--such as those caused by the transiting motion of a single outflowing cloud--it would be highly improbable for the absorption profile to return so closely to its initial state. 
A similar argument applies to models involving changes in covering fraction alone; such scenarios are unlikely to reproduce the observed profile stability across multiple epochs. 
These observations again support our interpretation that the variability in flow-I is governed by changes in the shielding gas column density from flow-II rather than by intrinsic changes in the properties of a single outflowing component.
However, we acknowledge that some observations remain unexplained by this model. For example, a narrow absorption feature present in region B during epoch 1 disappears in later epochs, suggesting that additional processes, such as localized changes in density, ionization, or cloud structure, may also play a role in shaping the detailed absorption profile.

\item The strong anti-correlation between BAL variability and UV-band flux, as discussed in subsection~\ref{subsec:light_curve}, supports our shielding-based interpretation. 
Increased shielding reduces high-energy ionizing photons, lowering the ionization parameter and enhancing the \civ\ fraction. 
However, the photo-ionization modelling suggests that the relatively small variations seen in the optical light curves (Fig.~\ref{fig:lc}), which trace the low-energy UV continuum in the rest frame, are insufficient to explain the observed BAL changes \citep{Aromal2021}. 
This aligns with our model, where the shielding gas primarily affects the high-energy UV and X-ray continuum that is  responsible for BAL variability, while leaving the lower-energy UV flux relatively unchanged.

\item The coordinated variability (albeit with varying amplitudes) across the entire velocity range of 6,000 to 30,000 \kms\ suggests a global mechanism influencing flow-I across possibly spatially separated regions of the outflow. This behavior is consistent with a shielding scenario, where changes in the shielding column can simultaneously affect multiple components of the flow.

\end{itemize}

Lastly, we estimate the typical transit timescale, $\Delta t_{\text{transit}} = D_{1500} / v_{\text{transverse}}$, where $D_{1500} \sim 0.008$ pc is the characteristic diameter of the continuum-emitting region at 1500\AA\ \citep[see][for detailed discussion]{Capellupo2013iii}, and $v_{\text{transverse}} \sim 20,000$ \kms\ is the assumed transverse velocity of the outflow across the line of sight, taken to be comparable to the average outflow velocity observed in \J11. This gives a transit timescale of $\Delta t_{\text{transit}} \sim 0.4$ yr, which is well within the observed variability timescale of $\Delta t \sim 3$ yr—i.e., the time between epochs 1 and 2, and also between epochs 2 and 3. 
Hence, these results are consistent with a scenario in which a flow component moves into our line of sight and subsequently exits, producing the observed emergence and disappearance of strong absorption features.

\subsection{Photo-ionization modeling in the presence of a shielding flow}

Here we consider our model in subsection~\ref{subsec:discussion1} and carry out the photo-ionization modeling using {\sc Cloudy} v23.01 \citep{cloudy2017} simulations.
We assume the outflow to be a plane parallel slab with solar metallicity and uniform density illuminated by a quasar spectral energy distribution (SED).
Under equilibrium conditions, {\sc Cloudy} computes the temperature and ionization states of the absorbing gas.

We consider the mean quasar SED from \citet{richards2006}  \citep[see also][]{Aromal2021} which goes through flow-II, that provides shielding in our model, before reaching flow-I which is located further away.
We perform two {\sc Cloudy} runs to test our model.
In the first {\sc Cloudy} run, the quasar SED is incident on the flow-II located close to the disk (assuming a high ionization parameter, log(U)$\sim$ 2)
with the total hydrogen column density ($N_{H, shield}$) varying from log($N_{H, shield}$) = 22 - 24.5 in 0.2 dex steps.
Now, if we assume a hydrogen density log($n_H$) = 7, then flow-II will be placed at the distance of a few sub-parsecs ($r\sim$ 0.02 pc) given log(U)$\sim$ 2.
Note that we use this distance as it is similar to the distance estimate one obtains using Keplerian arguments assuming v=$10^4$ \kms\ and log(M$_{BH}$) = 8.92 as calculated for \J11.
The specific values of both log(U) and $n_H$ assumed here are not important for our analysis and chosen in this way so that the distance to the flow-II is in the range of sub-parsecs, i.e. relatively close to the disk. 
We note that as $N_{H, shield}$ approaches high values (log($N_{H, shield}$) $\sim$ 23.5), the resulting \civ\ column density in flow-II itself can produce strong absorption as seen in BALs.
Next, we obtain the transmitted continuum after passing through the shielding gas having different $N_H$.
A comparison between incident and transmitted SED considered here is shown in \citet{Aromal2021} (Figure 13 therein). 
The shielding can significantly suppress the flux in the extreme UV and X-ray regimes, thereby can substantially alter the ionization conditions in a more distant outflow, i.e., flow-I.

In the second set of  {\sc Cloudy} runs, we use these transmitted continua as ionizing radiation on an absorber located further away at a distance of a few parsecs, say $\sim$ 5 pc.
This corresponds to flow-I in our picture.
We modeled this cloud with a total hydrogen column density, log($N_H$) = 18 having a uniform density log($n_H$) = 7.
We obtain the column density of each ionized species of interest such as \civ, \siiv, \aliii\ and \mgii\ in flow-I. We call them $N_{\text{shielded,ion}}$.
We also note these column density values for a different scenario where flow-I is illuminated by a quasar SED without any shielding, i.e. without the presence of flow-II, but located at the same distance as before, say, $N_{\text{ion}}$.

\begin{figure}
    \centering
    \includegraphics[viewport=30 10 2050 680, width=\textwidth,clip=true]{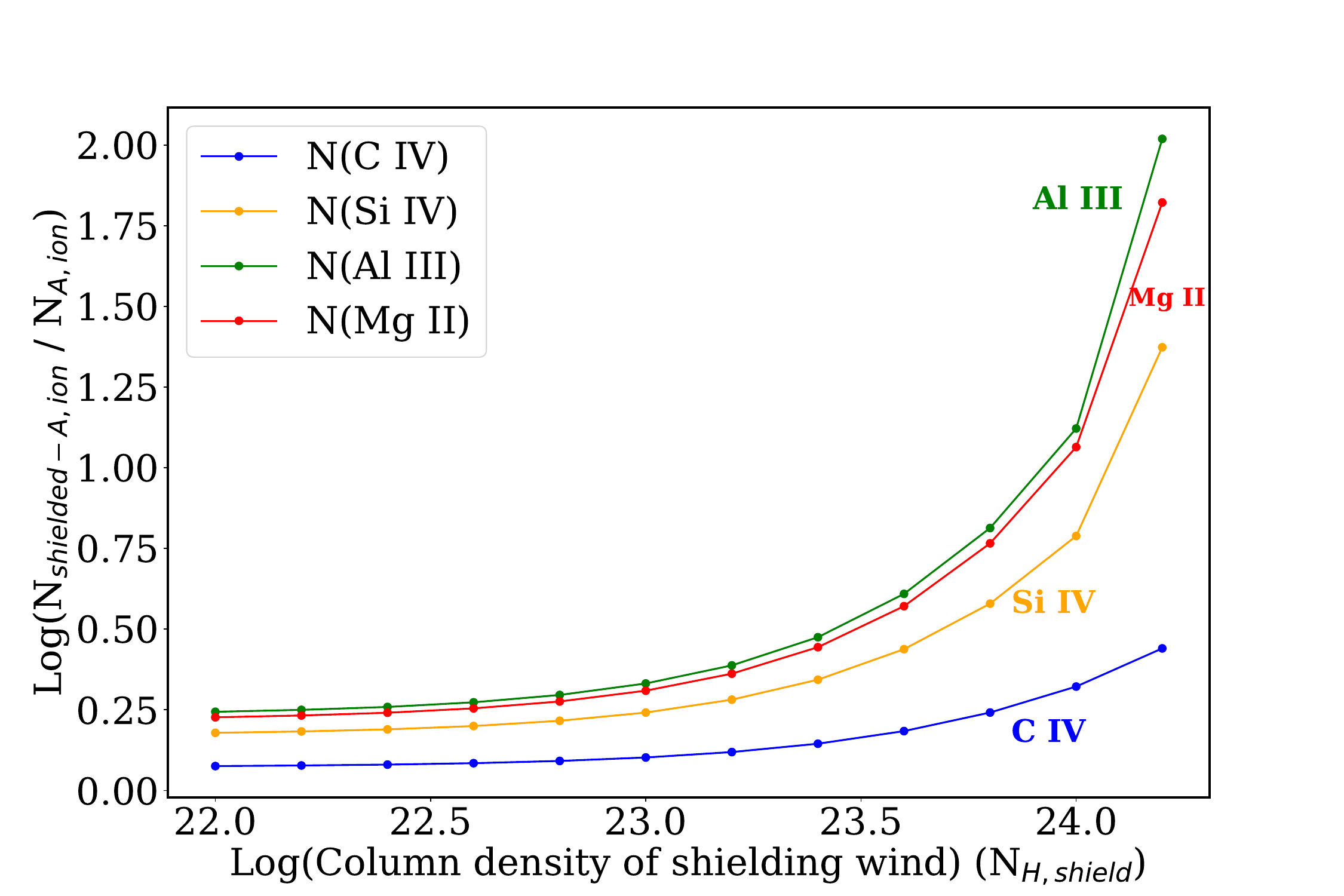}
    \caption{This figure shows the ratio of column density of different ionized species with and without the shielding gas ($N_{\text{shielded}, C IV}$ and $N_{C IV}$ respectively) as a function of the column density of shielding gas ($N_{H, shield}$). }
    \label{fig:cloudy_shielding_wind}
\end{figure}

In Fig~\ref{fig:cloudy_shielding_wind}, we plot $\frac{N_{\text{shielded,ion}}}{N_{ion}}$ as a function of the column density of shielding gas ($N_{H, shield}$).
This shows the effect of shielding gas in the ionic column densities of flow-I.
We see that $\frac{N_{shielded, C IV}}{N_{C IV}}$ does not change much with increasing $N_{H, shield}$ whereas low ionization species like \aliii\ and \mgii\ show more than an order of magnitude changes as log($N_{H, shield}$) approaches 24.
This indicates that the column density of low-ionization species is highly sensitive to the shielding gas in between.
This may explain the emergence of Lo-BAL absorption if an outflow is efficiently shielded by a high column density flow close to the disk.
Such a scenario is possible especially during the eruption of a strong wind close to the disk or during the transit of a strong flow component at relatively smaller distance as we propose in the case of \J11. 

This scenario is similar to the models presented in \citet{kaastra2014} and \citet{mehdipour2022}. 
For example, \citet{kaastra2014} observed the emergence of a relatively low-ionization, narrow UV absorption in a previously warm absorber like outflow in NGC 5548.
They concluded that this might be due to the presence of another outflow close to the accretion disk that provides large X-ray obscuration and at the same time produce BAL features in the UV spectra \citep[see Figure 4 in][]{kaastra2014}.
The obscuration decreases the ionization levels in the warm absorber further away which then appear as relatively low-ionization UV absorption.
To make it more interesting, \citet{mehdipour2022} reported that recent observations of NGC 5548 in 2022 indicated a significant reduction in X-ray obscuration and the near disappearance of narrow UV lines mentioned in \citet{kaastra2014}.
Additionally, \citet{fukumura2024ApJ} further demonstrated how accretion disk winds can result in both X-ray obscuration and UV absorption lines at the same time using MHD wind models.
This picture supports what we propose in the case of \J11 although we do not find any evidence of X-ray obscuration or X-ray absorption features in \J11 due to X-ray nondetection.

However, the question remains as to why only region D shows low-ionization absorption if the absorption from flow-I is present over a large velocity range.
This may be due to density variations along flow-I, as a change in U is highly dependent on the density of the cloud. In this case, region D may be probing a large region of high-density outflowing gas resulting in a large column density as observed and a relatively smaller U as well.
Although our model provides a highly likely scenario, we need more information to confirm the same.
Obtaining the distance to the outflows is a key for which we require an estimate of the density of the outflow through other methods such as using density sensitive transitions (e.g. \siv).
Also, a high-resolution spectrum may help us to resolve the narrow lines in the BAL region which may provide clues on the extent of saturation and possible kinematic shift in components seen, for example, in BAL region C.
Additionally, if our picture holds true, then such an analysis is essential for obtaining the covering fraction of the outflows which may help us to understand the extent to which multiple outflows can affect each other.



\section{Summary and Conclusions}
\label{sec:conclusions}


We present a detailed analysis of broad absorption line (BAL) variability in J115633.82+085628.9 using spectroscopic data from the Southern African Large Telescope (SALT) and SDSS, spanning nearly two decades across five epochs. 
This quasar is part of our high-velocity BAL sample of 63 quasars, which we have been monitoring with SALT for almost a decade \citep{aromal2023}. 
The BAL in this source spans a wide velocity range from 6,700 \kms to 30,000 \kms, although the upper limit is a conservative estimate and absorption can extend beyond 30,000 \kms.
For clarity, we divide the BAL into five regions, labeled A, B, C, D and E, corresponding to velocity ranges of 6,700–10,000 \kms, 10,000–12,500 \kms, 12,500–18,000 \kms, 18,000–22,000 \kms, and 22,000–30,000 \kms, respectively.

The \civ\ BAL exhibits extreme variability throughout our monitoring campaign, including one of the highest equivalent width variation rates ever reported in literature. 
Notably, we observe the emergence of low-ionization BALs (LoBALs) in \aliii\ and \mgii\ for the first time in epoch 2, coinciding with a significant increase in \civ\ absorption. 
These newly appeared LoBALs are found at high velocities (18,000–23,000 \kms) within region D, where the \civ\ absorption is also strongest—an unusual location for LoBALs, which are typically found at lower velocities.
Interestingly, by epoch 3, the \civ\ absorption weakens significantly, and the LoBALs completely disappear. 
{\it This is likely the first instance where the emergence and subsequent disappearance of LoBALs are observed within the same quasar, accompanied by such dramatic \civ\ changes.}

Regions A, B, and C also exhibit significant variability. Region A contains a strong, narrow \civ\ component with well-resolved \siiv\ absorption, both showing highly correlated variability. 
The narrow component in region B disappears completely after epoch 2 and does not reappear in subsequent observations. Region C, in contrast, hosts multiple narrow \civ\ components, displaying large velocity shifts indicative of kinematic variability. 
In region E, the \civ\ absorption depth increases dramatically ($\Delta\sim0.4$) in epoch 2 relative to epoch 1, while a \siiv\ absorption with flat-bottomed profile (indicating saturation) also emerges for the first time, with an average depth of $\sim0.4$.

Motivated by these observations, we propose a two flow model to explain the observed variability. In this picture, flow-I is a pre-existing outflow responsible for the absorption seen in epoch 1. 
In epoch 2, a new absorbing component, flow-II, with a large column density, emerges close to the accretion disk or transits into our line of sight. This additional component provides sufficient shielding to flow-I, leading to the temporary appearance of LoBAL features in region D. 
As flow-II eventually moves out of the line of sight, flow-I is once again exposed to the full ionizing radiation, resulting in the disappearance of the LoBALs. By epochs 4 and 5, the \civ\ BAL profile roughly returns to its original state from epoch 1. In particular, we did not observe a reappearance of the LoBALs in our subsequent monitoring with SALT.

We also performed simple photoionization modeling using {\sc Cloudy}, simulating a multi-flow scenario: one component close to the disk (corresponding to flow-II) and a second, more distant component (flow-I). 
Our models indicate that this configuration can provide sufficient shielding when the column density of flow-II increases and approaches the Compton thickness ($N_H \sim 23-24$). Under such conditions, the LoBAL column densities become highly sensitive to shielding, which leads to significant changes in optical depth and the emergence of LoBAL features, as observed in epoch 2.
There are precedents for this mechanism in the literature. For instance, in NGC 5548, the emergence of an X-ray absorbing BAL outflow was found to coincide with the appearance of new, relatively low-ionization UV absorption features in an already existing outflow component \citep{arav2015, kaastra2014, mehdipour2022}.
In summary, our results suggest that LoBALs sometimes can be transient features in quasars, triggered when substantial shielding is provided to pre-existing, distant parts of the outflow, rather than necessarily representing a specific evolutionary phase in quasar lifecycles.

\section*{Acknowledgements}
This work is supported by the Natural Science and Engineering Research Council (NSERC) RGPIN-2021- 04157 and a Western Research Leadership Chair Award. 
Based on observations collected at Southern African Large Telescope (SALT; Program IDs 2021-2-SCI-008, 2022-2-SCI-005 and 2023-2-SCI-002).

Funding for the Sloan Digital Sky Survey IV has been provided by the Alfred P. Sloan Foundation, the U.S. Department of Energy Office of Science, and the Participating Institutions. SDSS acknowledges support and resources from the Center for High-Performance Computing at the University of Utah. The SDSS web site is www.sdss4.org.

SDSS is managed by the Astrophysical Research Consortium for the Participating Institutions of the SDSS Collaboration including the Brazilian Participation Group, the Carnegie Institution for Science, Carnegie Mellon University, Center for Astrophysics | Harvard \& Smithsonian (CfA), the Chilean Participation Group, the French Participation Group, Instituto de Astrofísica de Canarias, The Johns Hopkins University, Kavli Institute for the Physics and Mathematics of the Universe (IPMU) / University of Tokyo, the Korean Participation Group, Lawrence Berkeley National Laboratory, Leibniz Institut für Astrophysik Potsdam (AIP), Max-Planck-Institut für Astronomie (MPIA Heidelberg), Max-Planck-Institut für Astrophysik (MPA Garching), Max-Planck-Institut für Extraterrestrische Physik (MPE), National Astronomical Observatories of China, New Mexico State University, New York University, University of Notre Dame, Observatório Nacional / MCTI, The Ohio State University, Pennsylvania State University, Shanghai Astronomical Observatory, United Kingdom Participation Group, Universidad Nacional Autónoma de México, University of Arizona, University of Colorado Boulder, University of Oxford, University of Portsmouth, University of Utah, University of Virginia, University of Washington, University of Wisconsin, Vanderbilt University, and Yale University.

\section*{Data Availability}
Data used in this work are obtained using SALT. Raw data will become available for public use 1.5 years after the observing date at https://ssda.saao.ac.za/.
\software{astropy \citep{astropy2022}, Cloudy \citep{cloudy2017}, {\sc PyQSOFit} \citep{guo2018a}}



\bibliography{mybib_bal}{}
\bibliographystyle{aasjournalv7}



\end{document}